\documentclass[12pt]{article}
\usepackage{amsmath, amssymb,graphicx, psfrag}

\newwrite\ffile\global\newcount\figno \global\figno=1

\def\writedef#1{}

\input epsf
\def\figin{\epsfcheck\figin}\def\figins{\epsfcheck\figins}
\def\epsfcheck{\ifx\epsfbox\UnDeFiNeD
\message{(NO epsf.tex, FIGURES WILL BE IGNORED)}
\gdef\figin##1{\vskip2in}\gdef\figins##1{\hskip.5in}% blank space instead
\else\message{(FIGURES WILL BE INCLUDED)}%
\gdef\figin##1{##1}\gdef\figins##1{##1}\fi}

\def\figinsert{}
\def\ifig#1#2#3{\xdef#1{fig.~\the\figno}
\writedef{#1\leftbracket fig.\noexpand~\the\figno}%
\figinsert\figin{\centerline{#3}}\medskip\centerline{\vbox{\baselineskip12pt
\advance\hsize by -1truein\center\footnotesize{  Fig.~\the\figno.} #2}}
\bigskip\endinsert\global\advance\figno by1}
\def\endinsert{}

\usepackage{graphics}

\begin{document}

\baselineskip 18pt
\newcommand{\Tr}{\mbox{Tr\,}}
\newcommand{\beq}{\begin{equation}}
\newcommand{\eeq}{\end{equation}}
\newcommand{\bea}{\begin{eqnarray}}
\newcommand{\eea}[1]{\label{#1}\end{eqnarray}}
\renewcommand{\Re}{\mbox{Re}\,}
\renewcommand{\Im}{\mbox{Im}\,}
\newcommand{\yms}{${YM^*\,}$}

\def\N{{\cal N}}

%%%%%%
%%%%%%%%%%%%%%%%   TITLE    %%%%%%%%%%%%%%%%%%%%

\thispagestyle{empty}
\renewcommand{\thefootnote}{\fnsymbol{footnote}}

{\hfill \parbox{4cm}{
        SHEP-06-11 \\
}}

\bigskip

\begin{center} \noindent \Large \bf
Improving the Infra-red of Holographic \\ Descriptions of QCD
\end{center}

\bigskip\bigskip\bigskip

\centerline{ \normalsize \bf Nick Evans, Andrew Tedder, Tom
Waterson \footnote[1]{\noindent \tt
 evans@phys.soton.ac.uk, ajmt@phys.soton.ac.uk, trw@phys.soton.ac.uk} }

\bigskip
\bigskip\bigskip

\centerline{ \it School of Physics and Astronomy} \centerline{ \it
Southampton University} \centerline{\it  Southampton, SO17 1BJ }
\centerline{ \it United Kingdom}
\bigskip

\bigskip\bigskip

\renewcommand{\thefootnote}{\arabic{footnote}}

\centerline{\bf \small Abstract} A surprisingly good holographic
description of QCD can be obtained from naive five dimensional
gauge theory on a truncated AdS space. We seek to improve the
infra-red description of QCD in such models by using a more
sophisticated metric and an action derived from string theory
duals of chiral symmetry breaking. Our metric is smooth into the
infra-red and the chiral condensate is a prediction of the
dynamics. The theory reproduces QCD meson data at the 10\% level.

\newpage

%%%%%%%%%%%%%%%%%%%%%%%%%%%%%%%%%%%%%%%%%%%%%%%%%%%%%%%%%

\section{Introduction}

The deep connections between QCD and string theory have been
revived in recent years by the AdS/CFT Correspondence
\cite{hep-th/9711200, hep-th/9802109, hep-th/9802150}. The
Correspondence provides an explicit description of a strongly
coupled gauge theory in terms of a weakly coupled,
holographic string description. The original conjecture was for a
highly supersymmetric conformal theory with only adjoint fields.
Technology has since been introduced that allows supersymmetry to
be broken and a running gauge coupling to be present
\cite{Girardello:1998pd,Girardello:1999bd,Freedman:1999gk}, and
the introduction of quark fields
\cite{hep-th/0205236,Bertolini:2001qa,Grana:2001xn}. Confinement
\cite{Maldacena:1998im,Rey:1998ik} and chiral symmetry breaking
\cite{hep-th/0306018}-\cite{Sakai:2005yt} have been investigated
using the string description.

Recently, the first attempts have been made to construct
phenomenological holographic models of QCD \cite{hep-ph/0501128,
hep-ph/0501218} (see also
\cite{Ghoroku:2005kg}-\cite{Karch:2006pv}). Surprisingly simple
models consisting of gauge theory in an anti-de-Sitter space
interval have turned out to provide a remarkably good description
of the meson sector of QCD. These models are in many ways naive
though. Amongst the criticisms that might be aimed at these models
are:
\medskip
\begin{itemize}
\item The use of an AdS geometry implicitly means that the
background gauge configuration is conformal (and essentially that
of large N ${\cal N}=4$ super Yang Mills).

\item The existence of a mass gap is imposed by hand through the
inclusion of a boundary to the space and is not the product of a
running coupling.

\item The fields that holographically describe the quark bilinears
are included phenomenologically and there is no rigorous (string
theory) realization of the construction.

\item The solution for the field which describes the quark mass
and condensate is also included by hand and the quark condensate
is not dynamically determined in terms of either the gauge
configuration or the quark mass.

\item The ultra-violet of the theory does not become
asymptotically free.

\item The excited meson mass spectrum typically scales like the
excitation number $n$ as opposed to the $\sqrt{n}$ scaling
predicted by a simple flux tube model \cite{hep-ph/0507246}.

\end{itemize}

In spite of these objections, the models do provide a good
description of the light meson sector of QCD. The clear next step
is to try to alleviate some or all of these objections. In this
paper we will address this task (progress has already been made in
\cite{Ghoroku:2005vt,Karch:2006pv}).

Our main tool will be to use the more rigorous AdS/CFT description
of chiral symmetry breaking in \cite{hep-th/0306018}. Previously
it has been used as a testing ground for the generic features of
chiral symmetry breaking \cite{hep-th/0403279}, but here we will
massage it to a phenomenological five dimensional holographic
description of QCD.

The geometry we will use is that on the surface of a D7 brane in a
non-supersymmetric dilaton flow deformation of the AdS/CFT
Correspondence. We review its origin in more detail in the
appendix, but let us stress its benefits now

\begin{itemize}
\item The background gauge configuration in which the quarks live
is non-supersymmetric (although not purely that of QCD) and has a
running coupling.

\item The mass gap is a result of the non-supersymmetric gauge
configuration and the geometry relevant for quark physics is
smooth at all radii or energy scales.

\item The holographic dual of the quark bilinear is explicit in
the string construction.

\item The quark condensate is a prediction of the gauge
configuration and is determined as a function of the quark mass.

\end{itemize}

These points go a considerable way towards addressing the
inconsistencies of the first models. We will, however, continue to
adopt the phenomenological approach with regards treating the
background as describing an N=3 rather than N$\rightarrow \infty$
theory. In addition, the string theory construction can only
realize a U(1) axial symmetry, and does not provide a holographic
dual of the axial vector mesons. We include by hand appropriate
fields to provide a non-abelian chiral symmetry and the axial
vector states in the phenomenological spirit of
\cite{hep-ph/0501128, hep-ph/0501218}.

One knows that the transverse parts of the vector vector and axial
axial correlators in QCD interact differently with the chiral
condensate in QCD. In the gravity dual one would expect the axial
and vector gauge fields to in fact see distinct metrics. We can
not incorporate this effect because the string model does not
provide enough information. Nevertheless the model links the quark
condensate to the dynamics and smooths the infra-red which should
improve the description, at least in the vector sector, whilst
doing no more harm in the axial sector than is done in
\cite{hep-ph/0501128, hep-ph/0501218}.

In this paper we compute with our phenomenological model the
masses and decay constants for the pion and the rho and $a_1$
vector mesons, and also the $g_{\rho \pi \pi}$ coupling. We find
that the model gives comparable predictions to the pure AdS models
within 12\% of the QCD values. We believe these results provide
support for the robustness of the predictions of these holographic
models.

The geometry we propose returns to pure AdS space in the
ultra-violet, so we do not address here the absence of asymptotic
freedom in the gravity description.  As we pointed out recently in
\cite{hep-th/0505250}, the gravity theory should only be used up
to a UV cut off, corresponding to the scale at which QCD switches
from perturbative to non-perturbative behaviour. Above that cut
off the gravitational dynamics must become non-perturbative with
its loop corrections completely dominating the classical results.
The correct UV dynamics should be encoded at that cut off by
correcting the values of higher dimension operator couplings. In
principle, these can be tuned in the AdS/CFT approach to produce
the holographic equivalent of a perfect lattice action.

As a small example of these ideas we consider the matching of the
five dimensional gauge coupling in the UV. In
\cite{hep-ph/0501128, hep-ph/0501218} this coupling is matched to
the perturbative result for the vector vector correlator in QCD.
The AdS gravitational dual presumably describes a strongly coupled
conformal theory in the UV and so the correlator behaviour matches
the logarithmic result of the conformal but weakly coupled UV
behaviour of QCD. It is surprising that the numerical coefficient
of the log term can be matched though. Here we test how good that
matching is by allowing the parameter to float and fitting it to
data. We find such a fit induces roughly a 30\% change in the
coupling value, which provides a measure of non-perturbative
corrections at the scale of matching to the strongly coupled
regime of QCD. We leave attempts to further improve the UV of the
theory for later work though.

Finally, it has recently been pointed out \cite{Karch:2006pv} that
an appropriate change to the IR behaviour of the dilaton can
correct the $n$ scaling of the tower of excited $\rho$ meson
states. We have tested our model in this respect but find only a
marginal improvement over the pure AdS case. This is a sign that,
although our geometry describes a non-supersymmetric gauge
configuration, it is still not a perfect description of QCD and
work remains to be done on improving the geometric background.

\section{Phenomenological Five Dimensional Models}

The phenomenological approaches to describing QCD holographically
are based on a 5d action of the form

\beq\label{eq:startaction}
  S \sim \int d^4x~ dr ~ e^{\phi} \sqrt{-g} \left(
  {\cal L}_\sigma + \sigma^2 Tr |D U|^2 - \frac{1}{4 g_5^2}~Tr (
  F_L^2 + F_R^2) \right)
\eeq where $D_\mu U=\partial_\mu U-iA_{L\mu}U+iUA_{R\mu}$. The
field $U(x,r)=\exp(i \pi^a(x,r)T^a)$ describes the pions produced
by the breaking of a $SU(N_f)$ chiral symmetry with generators
$T^a$. We assume that the background value of $U$ is the identity
so we are studying $N_f$ degenerate quarks. The non-abelian gauge
fields $A_L$ and $A_R$ couple by left and right action on $U$.
They will holographically describe the vector and axial vector
mesons. The field $\sigma$ is a function of $r$ only and
holographically describes the quark mass and $\langle \bar{q} q
\rangle$ expectation value.  A non-zero value for this field will
break the $SU(N_f)_L\times SU(N_f)_R$ chiral symmetry of the
action down to the vector $SU(N_f)_V$.

\subsection{Pure AdS}

In the simplest approaches \cite{hep-ph/0501128,hep-ph/0501218},
the dilaton, $\phi$, is taken to be constant, so drops from the
action. The background metric is AdS down to some boundary at
$r_0$ which breaks the conformal symmetry and provides the theory
with a mass gap. \beq
  ds^2 = \frac{r^2}{R^2} dx^2 + \frac{R^2}{r^2} dr^2,~~r_0\leq r<\infty.
\eeq Note that dilatation transformations in the field theory,
which define the mass dimension of operators (for example if we
scale $x \rightarrow e^\alpha x$ then a scalar field of dimension
one scales as $\phi \rightarrow e^{-\alpha} \phi$), are mapped to
a symmetry of the metric with the radial direction scaling as an
energy scale.

The Lagrangian for $\sigma$ in these models is given by \beq
  {\cal L}_\sigma = (\partial_r \sigma)^2 - 3 \sigma^2,
\eeq with resulting solutions $\sigma(r) = m/r + c/r^3$. Here
$\sigma$ has does not transform under the field theory dilatations
so $m$ has dimension one and $c$ dimension three. The two
parameters $m,c$ are fitted phenomenologically to the (degenerate)
light quarks' mass and condensate.

The remaining parameter is $g_5$, which in string theory duals is a
prediction in terms of the gauge theory 'tHooft coupling $g_{YM}^2
N$. In the phenomenological approach though, this relation is
abandoned and the value of $g_5$ is fitted to the vector current
correlator extracted from QCD.
\beq
  \int d^4x e^{iqx}\langle J^a_\mu(x)J^b_\nu(0)\rangle=\delta^{ab}
  (q_\mu q_\nu -q g_{\mu\nu})\Pi_V(-q^2),
\eeq
where $J^a_\mu(x)=\bar q \gamma_\mu T^a q$. For QCD, the leading order
contribution to $\Pi_V(-q^2)$ is \cite{Shifman:1978bx}
\beq\label{eq:piv-qcd}
  \Pi_V(-q^2)=-\frac{N}{24\pi^2}\ln(-q^2).
\eeq
In order to calculate this quantity from the five dimensional model,
we appeal to the AdS/CFT correspondence. The five dimensional vector
field $V^a_\mu(x,r)= (A^a_{L\mu}(x,r)+A^a_{R\mu}(x,r))$ acts as a source
for the four dimensional vector current $J^a_\mu(x)$ in the limit
$r\rightarrow\infty$. It obeys the equation of motion
\beq\label{eq:vec-eom}
  \partial_\mu\left(\frac{1}{g_5^2}~e^\phi\sqrt{-g}
  g^{\mu\alpha}g^{\nu\beta}(\partial_\alpha V^a_\beta
  - \partial_\beta V^a_\alpha)\right)=0.
\eeq
We look for solutions of the form $V^\mu(x,r)=V_0^\mu(x)v(x,r)$,
with $\lim_{r\rightarrow\infty}v(x,r)=1$, so that $V^\mu_0(x)$ will
act as a dimension one source for $J^a_\mu(x)$. Solving the equation
of motion (\ref{eq:vec-eom}) in the $V^r(x,r)=0$ gauge gives
\beq
  v(q,r)=-\frac{\pi}{2}\mathcal{Y}_1(q/r)\sim 1-\frac{q^2}{4r^2}\ln
  \left(\frac{-q^2}{r^2}\right), {\textrm as}~ r\rightarrow\infty,
\eeq
where $\mathcal{Y}_1$ is a Bessel function of the second kind.
Substituting the solution back into the action and differentiating
twice with respect to the source $V^\mu_0$ gives the vector current
correlator
\beq\label{eq:piv-ads}
  \Pi_V(-q^2)=\left[\frac{1}{g_5^2q^2}r^3\partial_rv(q,r)\right]_{r=\infty},
\eeq
which (up to contact terms) yields
\beq
  \Pi_V(-q^2)=-\frac{1}{2g_5^2}\ln(-q^2).
\eeq Finally, comparing this to the perturbative QCD result
(\ref{eq:piv-qcd}) determines the 5d coupling as \beq
\label{match}
  g_5^2 = \frac{12 \pi^2}{N}.
\eeq In \cite{hep-ph/0501218, hep-ph/0501128} this model is used
to calculate meson masses, decay constants and couplings
coefficients with great success. We summarize these results in
Table 1.

The matching in (\ref{match}) is of course naive. One should match
the gravitational theory to QCD only at the point where the QCD
coupling becomes non-perturbative where gluonic corrections to the
perturbative QCD result become important. It is therefore
interesting to recompute the results of \cite{hep-ph/0501218}, but
with $g_5$ being a free parameter of the model in order to see how
accurate this matching is. On performing a global fit on all of
the parameters, we found that the optimal value for $g_5$ is
$5.19$ which is $17\%$ smaller than the result $\sqrt{(12 \pi^2)/N}$
from matching to perturbative QCD. We conclude that
non-perturbative effects could have a significant effect.

\subsection{The New Model}

Our approach in this paper will be based around the D3/D7 brane
string theory construction described in the appendix
\cite{hep-th/0306018}. Here we will present the model as a 5d
model in the spirit of (\ref{eq:startaction}). Starting with the
string theory model's action (\ref{eq:5daction}), we construct a
phenomenological model by artificially extending the symmetry
group from $SU(N_f)_V \times U(1)_A$ to the chiral $SU(N_f)_L\times SU(N_f)_R$ and
add in the axial vector gauge field in (\ref{eq:startaction}).

The model has the metric \beq
  ds^2=H^{-1/2}f^{-\delta/4}\sum_{i=0}^3dx_i^2+H^{1/2}f^{1/2-\delta/4}h~dr^2,
\eeq
where
\begin{equation*}
  f=\frac{(\sigma(r)^2+r^2)^2+b^4}{(\sigma(r)^2+r^2)^2-b^4},~~
  h=\frac{(\sigma(r)^2+r^2)^2-b^4}{(\sigma(r)^2+r^2)^2},~~H=f^\delta-1,
\end{equation*}
and a radially changing dilaton and 5d gauge coupling
\begin{eqnarray}
  e^\phi=H^{5/4}f^{5/4-5\delta/8+\Delta/2}h^{5/2}r^3
  (1+\dot\sigma^2)^{-1/2}\sim r^{-2}, {\textrm as}~r\rightarrow\infty,\nonumber\\
  g_5^2=\hat g_5^2H^{1/2}f^{1/2-\delta/4+\Delta/2}h(1+\dot\sigma^2)^{-1}
  \sim \hat g_5^2 r^{-2}, {\textrm as}~r\rightarrow\infty.
\end{eqnarray}
with $\delta=1/2$, $\Delta=\sqrt{39}/2$. Note that we scaled all
coordinates by a factor of $R$. The conformal symmetry breaking
scale is fixed by the parameter $b$ which will determine the scale
$\Lambda_{QCD}$. Since it is the only scale in the model we set it
to one for computations.  At the string theory level the value of
$R$ fixes the 5d gauge coupling, but here we will fix that
phenomenologically to describe an $N_c=3$ theory so we have also
set $R=1$ and left $\hat{g}_5$ free. As $r\rightarrow\infty$, the
metric returns to $AdS_5$, the factor $e^\phi/g_5^2$ goes to
$1/\hat g_5^2$ and we are left with exactly the pure AdS model.
The radial dependence of the dilaton shows the model has a running
coupling \footnote{Formally in the string theory model of the
appendix the running coupling may be determined by placing a D3
brane probe in the geometry as in \cite{Buchel:2000cn} which gives
$g_{YM}^2 \sim (w^4 + b^4)/(w^4-b^4)^\Delta$. Note this running in
the strong coupling regime is not logarithmic and the gauge
coupling diverges at the scale $b$.}
\medskip

\noindent\emph{Dynamical Quark Condensate}

The chiral symmetry breaking quark condensate is determined
dynamically in this model by the background metric which
represents the background gauge configuration. The Lagrangian for
the field $\sigma(r)$  in this model is

\beq
  \mathcal{L}_\sigma=\sqrt{-g}f^{\Delta/2}g_{rr}^{3/2}\sqrt{1+\dot\sigma^2},
\eeq where the dot indicates differentiation with respect to $r$.
The equation of motion for this field, which is complicated since
$\sigma$ occurs throughout the geometry, is given by

\beq \label{eommc}
  { d \over d r} \left[ {f^{\Delta/2}  { \cal
  G}(r,\sigma) \over \sqrt{ 1 + (\partial_r\sigma)^2}  }
  (\partial_r \sigma)\right] - \sqrt{ 1 + \partial_r \sigma^2}
  { d \over d \bar{\sigma}} \left[ f^{\Delta/2} { \cal G}(r,\sigma)
  \right] = 0,
\eeq where \beq
  {\cal G}(r,\sigma) =  r^3 {( (r^2 + \sigma^{2})^2 +
  1) ( (r^2 + \sigma^{2})^2 - 1) \over (r^2 + \sigma^{2})^4}.
\eeq

\begin{figure}[!h]
\begin{center}
  \includegraphics[height=6cm,clip=true,keepaspectratio=true]{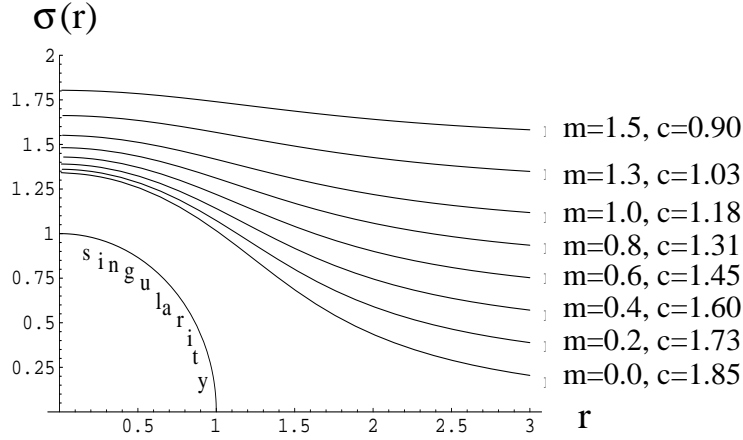}
  \caption{A plot of the embedding of the D7 brane as a function of the
  radial coordinate $r$}\label{f:embed}
\end{center}
\end{figure}

The large $r$ form of the solutions is of the AdS form (note from
the metric that $\sigma$ here enters symmetrically with $r$ and
therefore is rescaled relative to (3) and has energy dimension
one)

\beq \sigma(r) = m + c/r^2+\dots \eeq where $m$ and $c$ are
interpreted as the the quark mass and condensate respectively. We
seek regular solutions that satisfy $\dot\sigma(0)=0$. There is a
single such solution for each value of $\sigma(0)$ indicating that
the condensate $c$ is determined for a fixed asymptotic value of
$m$. The solutions are shown in Figure \ref{f:embed}.

Note that when the dynamical function $\sigma(r)$ is included in
the metric for the model there is no singularity since one cannot
reach $r+ \sigma=b$. The model therefore extends smoothly down to
$r=0$. We do not need to impose a hard IR cut off and the
conformal symmetry breaking is expressed through the parameter $b$
only.

\medskip
\noindent\emph{Matching the 5d Coupling}

The matching occurs at the boundary $r\rightarrow\infty$, so the results are exactly the same as those for the pure AdS calculation, and we are lead to the identification $\hat g_5^2=(12\pi^2)/N$.

\medskip
\noindent\emph{Vector Mesons}

We look for solutions to the vector equation of motion (\ref{eq:vec-eom}) that are of the form $V^a_\mu(x,r)=V^a_\mu(r)\exp(iqx)$. In the $V^a_r(x,r)=0$ gauge this gives the following equation of motion
\beq\label{eq:vec-eom-2}
   \partial_r(K_1(r)\partial_r V^a_\mu(r))+q^2K_2(r)V^a_\mu(r)=0,
\eeq
with
\begin{equation*}
K_1=f^{1/2}hr^3(1+\dot\sigma^2)^{-1/2},K_2=Hf^{1-\delta/2}h^2r^3(1+\dot\sigma^2)^{-1/2}.
\end{equation*}
We will interpret the rho mesons as normalisable modes of this equation, with the eigenvalues corresponding to the squared rho masses $m_\rho^2=-q^2$. For these modes to be normalisable, we require that they vanish sufficiently rapidly as $r\rightarrow\infty$. We must also impose the gauge invariant boundary condition $\psi_\rho'(0)=0$ to ensure the smoothness of the solution.

The rho wavefunction $\psi_\rho(r)$ is then a solution to (\ref{eq:vec-eom-2}) for an arbitrary component of $V^a_\mu(r)$ subject to the boundary conditions $\lim_{r\rightarrow\infty}\psi_\rho(r)=0$ and $\psi_\rho'(0)=0$. We solve the equation numerically to find the spectrum of rho masses.

For large $N$, one can write the vector current correlator as the sum over rho resonances
\beq\label{eq:piv-res}
  \Pi_V(-q^2)=-\sum_\rho\frac{F_\rho^2}{(q^2-m_\rho^2)m_\rho^2},
\eeq
where $F_\rho$ is the rho decay constant defined by $\langle 0|J^a_\mu|\rho^b\rangle =F_\rho~\delta^{ab}\epsilon_\mu$. In order to find $F_\rho$, we proceed by finding the Green's function solution to (\ref{eq:vec-eom-2}). Imposing the completeness relation
\beq
  \sum_\rho K_2(r)\psi_\rho(r)\psi_\rho(r')=\delta(r-r')
\eeq
on the set of eigenfunctions one finds
\beq
  G(q;r,r')=\sum_\rho\frac{\psi_\rho(r)\psi_\rho(r')}{q^2-m_\rho^2}.
\eeq
Generalising (\ref{eq:piv-ads}) we have
\beq
  \Pi_V(-q^2)=\left[\frac{1}{\hat g_5^2q^2} K_1(r)\partial_rv(q,r)\right]_{r=\infty}.
\eeq
It can be shown that, in terms of the Green's function, $v(q,r')=\left[K_1(r)\partial_rG(q;r,r')\right]_{r=\infty}$. From this, one finds
\beq
  \Pi_V(-q^2)=-\frac{1}{\hat g_5^2}\lim_{r\rightarrow\infty}\sum_\rho\frac{(K_1(r)\psi_\rho'(r))^2}{(q^2-m_\rho^2)m_\rho^2}.
\eeq
Comparing this to (\ref{eq:piv-res}) we can extract the rho decay constant
\beq\label{eq:frho}
  F_\rho^2=\frac{1}{\hat g_5^2}\lim_{r\rightarrow\infty}\left(K_1(r)\psi_\rho'(r)\right)^2.
\eeq

\medskip
\noindent\emph{The Axial Vector Mesons}

We write the axial vector field,
$A^a_\mu=(A^a_{L\mu}-A^a_{R\mu})$, in the $A_{a r}(x,r)=0$ gauge,
as perpendicular components plus a longitudinal component
$A^\mu_a= A^\mu_{a \bot} + \partial^\mu \phi$. The equation of
motion for the perpendicular components $A^a_{\mu \bot}$ with
$A_{i\bot}(x,r)=A(q,r)\exp(iqx)$ is \beq
  \partial_r(K_1(r)\partial_r A^a_\mu(r))+q^2K_2(r)A^a_\mu(r)-\hat g_5^2
  \sigma(r)^2K_3(r)A^a_\mu(r)=0,
\eeq
where $K_1(r)$ and $K_2(r)$ are the same as in (\ref{eq:vec-eom-2}), and
$K_3(r)=Hf^{3/2-\delta/2+\Delta/2}h^3r^3(1+\dot\sigma^2)^{-1/2}$.
The solutions represent the $a_1$ spin 1 axial vector meson if we
let $\lim_{r\rightarrow\infty}\psi_{a_1}(r)=0$, $\partial_r\psi_{a_1}(0)=0$.
We find the masses $m_{a_1}^2=-q^2$ by numerically finding the eigenvalues of
this equation. The decay constant $F_{a_1}$ is found in the same way as (\ref{eq:frho}).

The pion decay constant is similarly given by
 \beq
  f_\pi^2=\frac{1}{\hat g_5^2}\left[K_1(r)\partial_r \psi_{a_1}(0,r)\right]_{r=\infty}.
  \eeq
 We can then extract the quark mass using the
Gell-Mann-Oakes-Renner relation which must be obeyed for small
quark masses (ignoring the $m_q$ dependence of the condensate)\beq
  m_\pi^2f_\pi^2=2m_q c
\eeq

Results from these methods are displayed in table 1 and discussed
in section 3.\\

\noindent\emph{The Pion}

The pion and longitudinal axial gauge fields mix and one must look
for a solution of the coupled field equations

\beq \partial_r ( K_1(r) \partial_r \phi) + {\hat g_5}^2
\sigma(r)^2 K_3(r) (\pi^a - \phi^a) =0\eeq

\beq - q^2 K_1(r) \partial_r \phi + {\hat g_5}^2 K_4(r)
\sigma(r)^2 \partial_r \pi =0\eeq where
$K_4(r)=f^{1+\Delta/2}h^2r^3(1+\dot\sigma^2)^{-1/2}$.

The regular solutions of these equations require one to fix two
unknowns, the mass of the pion, $-q^2$, and the ratio of the
$\phi$ and $\pi$ fields at $r=0$. This is numerically hard.
Instead one can use the values of $m_\pi$ taken from the
Gell-Mann-Oakes-Renner relation and then find the ratio
$\phi(0)/\pi(0)$ with $\phi'(0)=0$ which leads to
$\phi(r=\infty)=\pi(r=\infty)=0$.

\noindent\emph{The Coupling $g_{\rho\pi\pi}$}

To the order we are working the value of the $g_{\rho\pi\pi}$
coupling can be read off from the expansion of $|DU|^2$ in the
action. This is not entirely satisfactory since $TrF^3$ terms,
which we don't include in the action, will also contribute.
Nevertheless for comparison to \cite{hep-ph/0501128} we will
compute them for our best fit models below. In particular

\beq \label{grpp} g_{\rho \pi \pi} =  \int dr {\hat g_5}
\psi_\rho(r) \left( {K_1(r) (\partial_r \phi)^2 \over {\hat
g_5}^2} + \sigma(r)^2 K_3(r) ( \pi^a - \phi^a) \right) \eeq

The $\pi$ field is normalized so the expression in brackets in
this last equation integrates to one.

\begin{table}
\begin{center}
  \begin{tabular}{|c|c|c|c|c|c|}\hline
  Observable       & Measured         & Model A   & AdS A      & Model B  & AdS B  \\
                   & (MeV)            & (MeV)     & (MeV)      & (MeV)    & (MeV)
                   \\\hline
  $m_\pi$          & $139.6\pm0.0004$ & $139.6^*$ & $139.6^*$  & $139.0$  & $141$  \\
   $m_\rho$        & $775.8\pm0.5$    & $775.8^*$ & $775.8^*$  & $742.7$  & $832$  \\
  $m_{a_1}$        & $1230\pm40$      & $1396$    & $1363$     & $1337$   & $1220$   \\
    $f_\pi$        & $92.4\pm0.35$    & $87.6$   & $92.4^*$   & $83.9$   & $84.0$  \\
  $F_\rho^{1/2}$   & $345\pm 8$        & $310.2$   & $329$      & $297.0$  & $353$  \\
  $F_{a_1}^{1/2}$  & $433\pm 13$       & $513.1$   & $486$      & $491.4$  & $440$  \\
  \hline
  \end{tabular}

\caption{ Results for meson variables in the models discussed in
the text. Model A is the new model in the paper with parameters
fixed to the starred measurements. AdS A is the equivalent pure
AdS model results with a hard IR cut off and the value of the
condensate being fitted.  Model B is a global fit in the new model
and AdS B is the equivalent fit result in pure AdS.}

\end{center}
\end{table}

\section{Results}

The results of the model are displayed in Table 1.  We compute 6
QCD meson parameters for our fits (we do not include $g_{\rho \pi
\pi}$). Our model has two free parameters (after fixing $g_5$
phenomenologically as discussed above), $b$ corresponding roughly
to the strong coupling scale $\Lambda$ and $m$ corresponding to
the light quark mass. The model therefore has the same number of
free parameters as real QCD.

In the first model, $A$, we match $b$ and $m$ by demanding that we
correctly reproduce $m_\pi$ and $m_\rho$. In order to do this, we
must set $\Lambda_b=264.5 ~MeV$ and $m=2.16~ MeV$. This gives a
prediction of $325.8~MeV$ for the scale of the quark condensate.
The overall rms error for this model is 12.8\% (Note
$\epsilon_{rms} = \sum_O ((\delta O/ O)^2/n)^{1/2}$ with $O$ the
observable and here $n=4$). For comparison we also reproduce the
pure AdS fit to the same parameters found in
\cite{hep-ph/0501128}. That model has three free parameters, the
value of the IR cut off, the quark mass and the quark condensate
and is therefore less predictive.

In model B, we perform a global fit to all observables. This gives
$\Lambda_b=253.2(MeV)$ and $m=2.24 ~MeV$, with the characteristic
scale for the quark condensate $311.9~MeV$. The overall rms error
for this model is 11.6\%. Again we reproduce the equivalent pure
AdS model fit for comparison.

For the best fit point we have also computed $g_{\rho \pi
\pi}=4.81 MeV$ using (\ref{grpp}). This should be compared to the
experimental result of $6.03\pm 0.07 MeV$ and to the results in
\cite{hep-ph/0501128} of   $4.48 MeV$ although, as discussed
above, the computation of this coupling is less robust than the
other results.

It is again interesting to test how well determined the 5d gauge
coupling $g_5$ is by the phenomenological fit to the far UV
expectation for $\Pi_{V}$. For example if one fits $\Lambda$,
$m_q$ and $g_5$ to correctly reproduce the three meson masses one
finds $g_5=4.36$ which is 30\% lower than the value
$\sqrt{12\pi^2/N_c}$ from perturbative QCD.

\section{Conclusions}

We have adapted a string theoretic model of chiral symmetry
breaking to a phenomenological description of QCD.  The model we
have proposed goes some way towards addressing the inconsistencies
of simple AdS slice holographic QCD models \cite{hep-ph/0501128,hep-ph/0501218}.
The background geometry of our model is non-supersymmetric, and it is the smooth
variation of this geometry with the radial direction $r$ that
provides a mass gap, without the need for an artificial hard IR
cut-off. In addition, the dual field to the quark mass/condensate
operator is a natural part of the geometrical set-up with the
value of the condensate being determined by the quark mass.

However, this is still a phenomenological approach in that we
introduce extra fields and symmetries by hand into the model in
order to describe the full pion and axial vector sectors. Formally
there is no geometric string interpretation for this system. We
also treat the background as though it describes an $N=3$ rather
than an $N=\infty$ field theory by matching the 5d gauge coupling
to QCD.

We find that the predictions of this model match experimental
results to within 12\%. This model is a little more predictive
than the pure AdS slice models since the condensate is dynamically
determined by the geometry. The best fit is in fact a few percent
worse than the AdS slice models but hopefully the theoretical
improvements represent at least a moral victory. In any case one
would naively have expected errors of order a few 100\% in all of
these models so the closeness to QCD across a range of holographic
models supports the robustness of the approach.

A drawback of these models to date has been that the geometry
returns to AdS for large $r$, meaning that the field theory is not
asymptotically free in the UV. Incorrect physics in the UV will
affect the strong coupling regime in the IR \cite{hep-th/0505250}.
Here we investigated corrections to the matching of the 5d gauge
coupling to naive perturbative QCD results. We found that this
coupling's value should be changed at the 30\% level indicating
the size of non-perturbative effects. In the future one might hope
to study the importance of higher dimension operators in the IR
physics as well.

\bigskip

\noindent {\bf \large Acknowledgements}

TW would like to thank PPARC for financial support.

\newpage

\renewcommand{\theequation}{A\arabic{equation}}
% redefine the command that creates the equation no.
\setcounter{equation}{0}  % reset counter
\section*{Appendix A - The String Theory Progenitor}  % use *-form to suppress numbering

The phenomenological model used here is based on the AdS/CFT
Correspondence realization of chiral symmetry breaking in \cite{hep-th/0403279}. That
model consists of a dilaton flow deformed AdS geometry

\beq ds^2 = H^{-1/2} \left( { w^4 + b^4 \over w^4-b^4}
\right)^{\delta/4} dx_{4}^2 + H^{1/2} \left( {w^4 + b^4 \over w^4-
b^4}\right)^{(2-\delta)/4} {w^4 - b^4 \over w^4 } \sum_{i=1}^6
dw_i^2, \eeq where

\beq H =  \left(  { w^4 + b^4 \over w^4 - b^4}\right)^{\delta} - 1
\eeq and the dilaton and four-form are given by

\beq e^{2 \phi} = e^{2 \phi_0} \left( { w^4 + b^4 \over w^4 - b^4}
\right)^{\Delta}, \hspace{1cm} C_{(4)} = - {1 \over 4} H^{-1} dt
\wedge dx \wedge dy \wedge dz. \eeq There are formally two free
parameters, $R$ and $b$, since \beq \delta = {R^{4} \over 2 b^4},
\hspace{1cm} \Delta^2 = 10 - \delta^2 \eeq

We can see that dimensionally $b$ has energy dimension one and
enters to the fourth power. The SO(6) symmetry of the geometry is
retained at all $r$. We conclude that in the field theory a
dimension four operator with no SO(6) charge has been switched on.
$b^4$ therefore corresponds to a vev for the operator Tr$F^2$.

Quarks are introduced by including probe D7 branes into the
geometry. As shown in figure 1, strings which stretch between the
D3 and D7 branes are in the fundamental representation of the
SU(N) gauge theory on the D3. The length of the minimum length
string between the two branes determines the mass of the quark
field. We minimize the D7's world-volume in the spacetime around
the D3 branes. This is encoded by the Dirac Born Infeld action in
Einstein frame of the D7 brane

\beq
  S_{D7}=-\tau_7\int d^8\xi~e^\phi\left[-\det({\bf P}[g_{ab}])\right]^{\frac{1}{2}},
\eeq
 where the pull back of the metric ${\bf P}[g_{ab}]$ is given by

\beq
  {\bf P}[g_{ab}] = g_{MN} {dx^M \over d \xi^a} {dx^N \over d \xi^b}
\eeq

\begin{figure}[!h]
\begin{center}
\begin{tabular}{cc}
\includegraphics[height=4cm,clip=true,keepaspectratio=true]{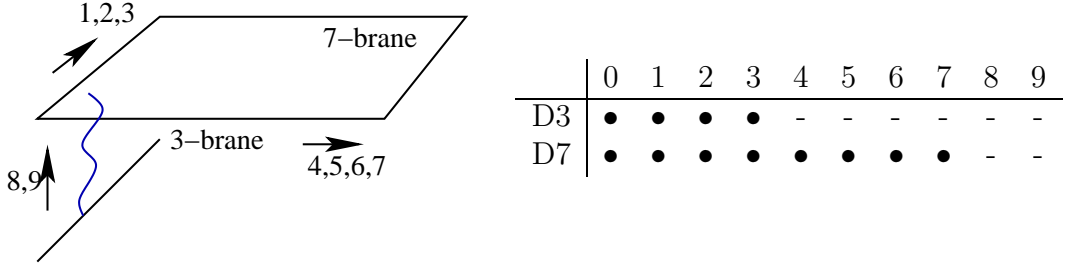}
\vspace{-3cm}& \\ & \begin{tabular}{c|cccccccccc} & 0 & 1 & 2 & 3
& 4 & 5 & 6 & 7 & 8 & 9 \\ \hline  D3 & $\bullet$ &$\bullet$
&$\bullet$ &$\bullet$ & -& -& -& -& -&
-\\
D7 & $\bullet$ &$\bullet$ &$\bullet$ &$\bullet$ & $\bullet$&
$\bullet$& $\bullet$& $\bullet$& -& -\end{tabular} \end{tabular}
\vspace{1.5cm}

 \caption{The basic
geometry of the D3 D7 system under consideration.}\label{fig1}
\end{center}
\end{figure}

Substituting from the geometry above we can find the equation of
motion for the radial separation, $\sigma$, of the two branes in
the $8,9$ directions as a function of the radial coordinate $r$ in
the $4-7$ directions. It is just eq.(\ref{eommc}) with the
solutions shown in Figure {\ref{f:embed}}. The solutions show that a dynamical mass
is formed for the quarks. A massless quark would correspond to a
D7 brane that intersects the D3 brane so there was a zero length
string between them. We see that the D3s repel the D7 and for all
configurations there is a minimum length string. The solution
which asymptotically has $m=0$ also explicitly break the U(1)
symmetry in the $8,9$ plane by bending off the axis. This is the
geometric representation of the breaking of the U(1) axial
symmetry of the quarks.

Fluctuations of the brane about the solution found above in the
$8,9$ directions correspond to excitations of the operator
$\bar{q} q$ and contain information about the pion and sigma field
of the model. Letting $u_8+iu_9 = \sigma(r)U(r,\xi)$ and expanding to
second order in $U(r,\xi)$ gives
\beq
  S=-\tau_7\int d^8\xi~e^\phi\sqrt{-g}(1+\dot\sigma^2)^{\frac{1}{2}}
  \left[1+\frac{1}{2} g_{rr}\sigma^2(1+\dot\sigma^2)^{-1}\partial^aU\partial_aU^\dag\right].
\eeq
Letting $U(r,\xi)=\exp(i\pi(r,\xi)$, this gives an action for the
pion field and the $\sigma$ field.

There is also a superpartner U(1) gauge field in the action which
describes the operator $\bar{q} \gamma^\mu q$ and hence vector
mesons. This is introduced as a gauge field $F_{ab}$ living on the
D7 \beq
  S=-\tau_7\int d^8\xi~e^\phi\left[-\det({\bf P}[g_{ab}]+2\pi\alpha'
  e^{-\phi/2}F_{ab})\right]^{\frac{1}{2}},
\eeq which, expanded to second order gives \beq
  S=-\tau_7\int d^8\xi~e^\phi\sqrt{-g}(1+\dot\sigma^2)^{\frac{1}{2}}
  \left[1+\frac{1}{2} g_{rr}{ \sigma^2 \over (1+\dot\sigma^2)}\partial^aU
  \partial_aU^\dag - \frac{1}{4}{(2\pi\alpha'^2)\over (1+\dot\sigma^2)}
  e^{-\phi}F^2\right].
\eeq Now, if we assume that the fields do not have any components
on the three sphere (which is appropriate for duals to
non-sypersymmetric fields) we arrive at the 5d action
\beq\label{eq:5daction}
  S = -R^{-8}\int d^4x~ dr ~ e^{\phi} \sqrt{-g} \left(
  {\cal L}_\sigma + \sigma^2  |\partial U|^2 - \frac{1}{4 g_5^2}~F^2) \right)
\eeq Between (A9) and (A10) we have rescaled $U\rightarrow
R^4\pi\sqrt{\tau_7}U$ and redefined the metric, dilaton and $g_5$
to the appropriate notation for a 5d model. In particular $g_5$
and the dilaton now have additional $r$ dependence which is just
that found in (12). The asymptotic large $r$ value of $g_5$ is
given by $16\pi^3g_s\alpha'^2/r^2$. Normalizable solutions of the
fields in this model in the fifth direction correspond to physical
states in the gauge theory, with the quantum numbers of the
operators described by the holographic field. Integrating over $r$
then leaves the four dimensional effective Lagrangian for these
states from which masses and couplings can be read off. A more
complete analysis of this model can be found in
\cite{hep-th/0306018,hep-th/0403279,Apreda:2005hj}. This method
can be extended to give a theory with an $SU(N_f)_V \times U(1)_A$
symmetry by replacing the single D7 brane with a stack of $N_f$ D7
branes (at large $r$ the theory becomes supersymmetric and there
is a superpotential term linking the adjoint matter fields and the
quarks which breaks the $SU(N_f)_A$ symmetry). We must, however,
be careful to keep $N_f << N$ so that we can still treat the stack
of D7 branes as a probe, and ignore any back reaction on the
geometry.


\begin{thebibliography}{ll}
\bibitem{hep-th/9711200}
  J.~M.~Maldacena,
  %``The large N limit of superconformal field theories and supergravity,''
  Adv.\ Theor.\ Math.\ Phys.\ {\bf 2} (1998) 231
  [Int.\ J.\ Theor.\ Phys.\ {\bf 38} (1999) 1113]
  [arXiv:hep-th/9711200].
  %%CITATION = HEP-TH 9711200;%%

\bibitem{hep-th/9802109}
  S.~S.~Gubser, I.~R.~Klebanov and A.~M.~Polyakov,
  %``Gauge theory correlators from non-critical string theory,''
  Phys.\ Lett.\ B {\bf 428} (1998) 105
  [arXiv:hep-th/9802109].
  %%CITATION = HEP-TH 9802109;%%

\bibitem{hep-th/9802150}
  E.~Witten,
  %``Anti-de Sitter space and holography,''
  Adv.\ Theor.\ Math.\ Phys.\ {\bf 2} (1998) 253
  [arXiv:hep-th/9802150].
  %%CITATION = HEP-TH 9802150;%%

  %\cite{Girardello:1998pd}
\bibitem{Girardello:1998pd}
  L.~Girardello, M.~Petrini, M.~Porrati and A.~Zaffaroni,
  %``Novel local CFT and exact results on perturbations of N = 4 super
  %Yang-Mills from AdS dynamics,''
  JHEP {\bf 9812} (1998) 022
  [arXiv:hep-th/9810126].
  %%CITATION = HEP-TH 9810126;%%

%\cite{Girardello:1999bd}
\bibitem{Girardello:1999bd}
  L.~Girardello, M.~Petrini, M.~Porrati and A.~Zaffaroni,
  %``The supergravity dual of N = 1 super Yang-Mills theory,''
  Nucl.\ Phys.\ B {\bf 569} (2000) 451
  [arXiv:hep-th/9909047].
  %%CITATION = HEP-TH 9909047;%%

%\cite{Freedman:1999gk}
\bibitem{Freedman:1999gk}
  D.~Z.~Freedman, S.~S.~Gubser, K.~Pilch and N.~P.~Warner,
  %``Continuous distributions of D3-branes and gauged supergravity,''
  JHEP {\bf 0007} (2000) 038
  [arXiv:hep-th/9906194].
  %%CITATION = HEP-TH 9906194;%%

  \bibitem{hep-th/0205236}
  A.~Karch and E.~Katz,
  %``Adding flavor to AdS/CFT,''
  JHEP {\bf 0206} (2002) 043
  [arXiv:hep-th/0205236].
  %%CITATION = HEP-TH 0205236;%%

  %\cite{Bertolini:2001qa}
\bibitem{Bertolini:2001qa}
M.~Bertolini, P.~Di Vecchia, M.~Frau, A.~Lerda and R.~Marotta,
%``N = 2 gauge theories on systems of fractional D3/D7 branes,''
Nucl.\ Phys.\ B {\bf 621} (2002) 157 [arXiv:hep-th/0107057].
%%CITATION = HEP-TH 0107057;%%

%\cite{Grana:2001xn}
\bibitem{Grana:2001xn}
M.~Grana and J.~Polchinski,
%``Gauge / gravity duals with holomorphic dilaton,''
Phys.\ Rev.\ D {\bf 65} (2002) 126005 [arXiv:hep-th/0106014].
%%CITATION = HEP-TH 0106014;%%


%\cite{Maldacena:1998im}
\bibitem{Maldacena:1998im}
  J.~M.~Maldacena,
  %``Wilson loops in large N field theories,''
  Phys.\ Rev.\ Lett.\  {\bf 80} (1998) 4859
  [arXiv:hep-th/9803002].
  %%CITATION = HEP-TH 9803002;%%

%\cite{Rey:1998ik}
\bibitem{Rey:1998ik}
S.~J.~Rey and J.~T.~Yee,
%``Macroscopic strings as heavy quarks in large N gauge theory and  anti-de
%Sitter supergravity,''
Eur.\ Phys.\ J.\ C {\bf 22} (2001) 379 [arXiv:hep-th/9803001].
%%CITATION = HEP-TH 9803001;%%


\bibitem{hep-th/0306018}
  J.~Babington, J.~Erdmenger, N.~J.~Evans, Z.~Guralnik and I.~Kirsch,
  %``Chiral symmetry breaking and pions in non-supersymmetric gauge / gravity
  %duals,''
  Phys.\ Rev.\ D {\bf 69} (2004) 066007
  [arXiv:hep-th/0306018].
  %%CITATION = HEP-TH 0306018;%%

  \bibitem{hep-th/0403279}
  N.~J.~Evans and J.~P.~Shock,
  %``Chiral dynamics from AdS space,''
  Phys.\ Rev.\ D {\bf 70} (2004) 046002
  [arXiv:hep-th/0403279].
  %%CITATION = HEP-TH 0403279;%%

  %\cite{Evans:2005ti}
\bibitem{Evans:2005ti}
N.~Evans, J.~Shock and T.~Waterson,
%``D7 brane embeddings and chiral symmetry breaking,''
JHEP {\bf 0503} (2005) 005 [arXiv:hep-th/0502091].
%%CITATION = HEP-TH 0502091;%%




%\cite{Ghoroku:2004sp}
\bibitem{Ghoroku:2004sp}
K.~Ghoroku and M.~Yahiro,
%``Chiral symmetry breaking driven by dilaton,''
Phys.\ Lett.\ B {\bf 604} (2004) 235 [arXiv:hep-th/0408040].
%%CITATION = HEP-TH 0408040;%%

%\cite{Kruczenski:2003uq}
\bibitem{Kruczenski:2003uq}
M.~Kruczenski, D.~Mateos, R.~C.~Myers and D.~J.~Winters,
%``Towards a holographic dual of large-N(c) QCD,''
JHEP {\bf 0405} (2004) 041 [arXiv:hep-th/0311270].
%%CITATION = HEP-TH 0311270;%%

%\cite{Bak:2004nt}
\bibitem{Bak:2004nt}
D.~Bak and H.~U.~Yee,
%``Separation of spontaneous chiral symmetry breaking and confinement via
%AdS/CFT correspondence,''
Phys.\ Rev.\ D {\bf 71} (2005) 046003 [arXiv:hep-th/0412170].
%%CITATION = HEP-TH 0412170;%%

%\cite{Sakai:2004cn}
\bibitem{Sakai:2004cn}
T.~Sakai and S.~Sugimoto,
%``Low energy hadron physics in holographic QCD,''
Prog.\ Theor.\ Phys.\  {\bf 113} (2005) 843
[arXiv:hep-th/0412141].
%%CITATION = HEP-TH 0412141;%%



%\cite{Sakai:2005yt}
\bibitem{Sakai:2005yt}
T.~Sakai and S.~Sugimoto,
%``More on a holographic dual of QCD,''
Prog.\ Theor.\ Phys.\  {\bf 114} (2006) 1083
[arXiv:hep-th/0507073].
%%CITATION = HEP-TH 0507073;%%

\bibitem{hep-ph/0501128}
  J.~Erlich, E.~Katz, D.~T.~Son and M.~A.~Stephanov,
  %``QCD and a holographic model of hadrons,''
  arXiv:hep-ph/0501128.
  %%CITATION = HEP-PH 0501128;%%

\bibitem{hep-ph/0501218}
  L.~Da Rold and A.~Pomarol,
  %``Chiral symmetry breaking from five dimensional spaces,''
  Nucl.\ Phys.\ B {\bf 721} (2005) 79
  [arXiv:hep-ph/0501218].
  %%CITATION = HEP-PH 0501218;%%

%\cite{Ghoroku:2005kg}
\bibitem{Ghoroku:2005kg}
K.~Ghoroku and M.~Yahiro,
%``Holographic model for mesons at finite temperature,''
arXiv:hep-ph/0512289.
%%CITATION = HEP-PH 0512289;%%

%\cite{Hambye:2005up}
\bibitem{Hambye:2005up}
T.~Hambye, B.~Hassanain, J.~March-Russell and M.~Schvellinger,
%``On the Delta(I) = 1/2 rule in holographic QCD,''
arXiv:hep-ph/0512089.
%%CITATION = HEP-PH 0512089;%%

%\cite{Ghoroku:2005vt}
\bibitem{Ghoroku:2005vt}
K.~Ghoroku, N.~Maru, M.~Tachibana and M.~Yahiro,
%``Holographic model for hadrons in deformed AdS(5) background,''
Phys.\ Lett.\ B {\bf 633} (2006) 602 [arXiv:hep-ph/0510334].
%%CITATION = HEP-PH 0510334;%%

%\cite{DaRold:2005vr}
\bibitem{DaRold:2005vr}
L.~Da Rold and A.~Pomarol,
%``The scalar and pseudoscalar sector in a five-dimensional approach to chiral
%symmetry breaking,''
JHEP {\bf 0601} (2006) 157 [arXiv:hep-ph/0510268].
%%CITATION = HEP-PH 0510268;%%

%\cite{deTeramond:2005su}
\bibitem{deTeramond:2005su}
G.~F.~de Teramond and S.~J.~Brodsky,
%``The hadronic spectrum of a holographic dual of QCD,''
Phys.\ Rev.\ Lett.\  {\bf 94} (2005) 201601
[arXiv:hep-th/0501022].
%%CITATION = HEP-TH 0501022;%%

%\cite{Hirn:2005nr}
\bibitem{Hirn:2005nr}
J.~Hirn and V.~Sanz,
%``Interpolating between low and high energy QCD via a 5D Yang-Mills model,''
JHEP {\bf 0512} (2005) 030 [arXiv:hep-ph/0507049].
%%CITATION = HEP-PH 0507049;%%



%\cite{Hirn:2005vk}
\bibitem{Hirn:2005vk}
J.~Hirn, N.~Rius and V.~Sanz,
%``Geometric approach to condensates in holographic QCD,''
arXiv:hep-ph/0512240.
%%CITATION = HEP-PH 0512240;%%




%\cite{Hong:2006si}
\bibitem{Hong:2006si}
D.~K.~Hong and H.~U.~Yee,
%``Holographic estimate of oblique corrections for technicolor,''
arXiv:hep-ph/0602177.
%%CITATION = HEP-PH 0602177;%%

%\cite{Shock:2006qy}
\bibitem{Shock:2006qy}
  J.~P.~Shock and F.~Wu,
  %``Three flavour QCD from the holographic principle,''
  arXiv:hep-ph/0603142.
  %%CITATION = HEP-PH 0603142;%%


%\cite{Karch:2006pv}
\bibitem{Karch:2006pv}
A.~Karch, E.~Katz, D.~T.~Son and M.~A.~Stephanov,
%``Linear Confinement and AdS/QCD,''
arXiv:hep-ph/0602229.
%%CITATION = HEP-PH 0602229;%%

\bibitem{hep-ph/0507246}
  M.~Shifman,
  %``Highly excited hadrons in QCD and beyond,''
  arXiv:hep-ph/0507246.
  %%CITATION = HEP-PH 0507246;%%


\bibitem{hep-th/0505250}
  N.~Evans, J.~P.~Shock and T.~Waterson,
  %``Towards a perfect QCD gravity dual,''
  Phys.\ Lett.\ B {\bf 622} (2005) 165
  [arXiv:hep-th/0505250].
  %%CITATION = HEP-TH 0505250;%%

\bibitem{Shifman:1978bx}
  M.~A.~Shifman, A.~I.~Vainshtein and V.~I.~Zakharov,
  %``QCD And Resonance Physics. Sum Rules,''
  Nucl.\ Phys.\ B {\bf 147} (1979) 385.
  %%CITATION = NUPHA,B147,385;%%



%\cite{Apreda:2005hj}
\bibitem{Apreda:2005hj}
  R.~Apreda, J.~Erdmenger and N.~Evans,
  %``Scalar effective potential for D7 brane probes which break chiral
  %symmetry,''
  arXiv:hep-th/0509219.
  %%CITATION = HEP-TH 0509219;%%

  %\cite{Buchel:2000cn}
\bibitem{Buchel:2000cn}
  A.~Buchel, A.~W.~Peet and J.~Polchinski,
  % ``Gauge dual and noncommutative extension of an N = 2 supergravity
  %solution,''
  Phys.\ Rev.\ D {\bf 63} (2001) 044009
  [arXiv:hep-th/0008076].
  %%CITATION = HEP-TH 0008076;%%




\end{thebibliography}
\end{document}